# Elastic fractal higher-order topological states

**Tingfeng Ma[*], Bowei Wu, Jiachao Xu, Hui Chen[†], Shuanghuizhi Li, Boyue Su, Pengfei Kang, Ji Wang**

Zhejiang-Italy Joint Lab for Smart Materials and Advanced Structures, School of Mechanical Engineering & Mechanics, Ningbo University, Ningbo, 315211, China

## ABSTRACT

Fractal is an intriguing geometry with self-similarity and non-integer dimensions, the elastic-wave topological phase based on fractal structures has not been revealed up to now. In this work, elastic-wave higher-order topological states in fractal structures are investigated. Elastic real-space quantized quadrupole moment is calculated and used to characterize the topology of elastic fractal metamaterials, and formation conditions of topological phase transitions in elastic fractal systems are revealed. The topological edge and corner states of elastic waves in fractal structures are realized theoretically and experimentally. It is found that different from the acoustic fractal system, the topological outer and inner edge states can emerge separately in elastic fractal systems, which is important for the integrated sensing and particle manipulation in microfluidics. Besides, the results show that the robustness of the topological corner states in rhombus fractal structures is obviously stronger than that in Sierpinski fractal structures, and the physical mechanism is clarified. Compared with traditional elastic-wave topological insulators based on periodic structures, the richness of topological states in elastic fractal structures is much higher (for the Sierpinski fractal structure, the number of topological states is 156, much greater than that of the periodic structure (only 28)), which is vital in integrated sensing and energy-location applications. The topological phenomena of elastic fractal systems revealed in this work, provides an unprecedented way of controlling elastic waves, enriches the topological physics of elastic systems and breaks the limitation of that relying on periodic elastic structures. The results have great application prospects in high-Q resonators, high-resolution elastic-wave energy locations, energy harvester, and high-sensitivity sensors.

[*]matingfeng@nbu.edu.cn,   [†]chenhui2@nbu.edu.cn

# I. Introduction

In recent years, topological insulators [1-3] have shown great potentials in wave-transmission controls, based on which, wave transmissions with extremely low loss and strong immunity to structural defects have been realized [4–9]. Topological insulators have been successfully applied in fields of electromagnetic [10–13], acoustic [14–17], elastic waves [18–26].

Topological transmissions of waves bring important opportunities for wave communication and sensing by improving signal-to-noise ratios, resolutions and sensitivities [27, 28]. Specially, Fan et al. [29] realized the higher-order topological states with high energy-concentration in a two-dimensional (2D) continuous elastic periodic system, experimentally observed the gapped one-dimensional (1D) edge states, and the topologically protected 0D corner state, which is vital to improve the sensitivity of elastic-wave sensors. Hong et al. [30] explored the valley-selectivity of corner states at high frequencies by engineering positions of the resonators, which provides generality for realizing high-frequency edge states and corner states. Liu et al. [31] investigates the higher-order topological behavior of elastic-wave metamaterials composed of L-shaped pillars attached to a plate, and realized physical transferrings among the corner, edge, and bulk elastic-wave modes by tuning the rotation angle of the L-shaped pillars.

For higher-order topological states, previous studies mostly focused on only systems in an integer-dimension [32–35], namely periodic systems satisfying the Bloch's theorem. In integrated sensing, particle manipulation and energy acquisition applications, a large number of topological states are required in a specific operating frequency range to adapt to changes in external load conditions. However, the number of topological states generated in periodic systems is very limited, which cannot meet the requirements on richness of topological states in those application fields. In recent years, rich topological states generated in fractal systems in a non-integer dimension have received much attentions.

Fractal [36] is a geometric form with self-similarity, and its unique structure and properties provide a new idea for metamaterial designs in the field of wave manipulations [37]. Based on fractal, many fascinating topological phenomena have been discovered [38–41]. Yang et al. [42] proved that photonic fractal Floquet topological insulators can be realized in photonic lattice with a Sierpinski carpet composed of spiral optical waveguides. Zheng et al. [43] et al. experimentally realized acoustic higher-order topological states in fractal dimensional systems, which supports both outer and inner corner states. Li et al. [44] observed the one-way edge states that are protected by a robust mobility gap and demonstrated the fundamental interplay between the fractality and topological Haldane insulator. They also experimentally realized higher-order topological insulator in an acoustical lattice and found that there exists a plenty of corner states [45].

However, the studies of higher-order topological phenomena based on fractal structures are currently limited to the fields of photonic [46–48] and acoustic waves [43, 45]. In elastic systems, rich topological edge and corner states are very important for realizing elastic-wave sensings, particle manipulations and resonating systems with high resolution and stability. Up to now, the topological states in elastic fractal systems have not been realized, which hinders the design of high-performance elastic wave devices.

In this work, the elastic fractal higher-order topological states are investigated. Elastic real-space quantized quadrupole moment is calculated and used to characterize the topology of elastic fractal systems, and the forming conditions of the topological phase transition in elastic fractal systems are clarified. Then, elastic higher-order topological states in the Sierpinski and rhombic fractal structures are realized numerically and experimentally, the strong robustness of which are further verified.

The structure of this paper is as follows: Section II presents the calculation of the elastic real-space quantized quadrupole moment and the characterization of topological phase transitions of elastic fractal systems. In Section III, the topological edge and corner states in Sierpinski and rhombic fractal structures are obtained by numerical simulations and experiments; the strong robustness of elastic topological states in fractal structures are verified by introducing defects and lattice disorders. The conclusion is given in Section IV.

## II. Model and theory

The propagations of elastic waves in the Sierpinski fractal structure shown in Fig.1(a) are considered. Through the box-counting method [49], the dimension number of this fractal model can be calculated: $d_f = \ln 8 / \ln 3 \approx 1.893$. The unit cell is shown in Fig.1(b) and (c), the unit constant is $a = 25$ mm, the plate thickness and column thickness are respectively $h_1 = 2$ mm and $h_2 = 6$ mm. The energy band of the elastic fractal structure can be well approximated by the elastic analog of the generalized 2D Su-Schrieffer-Heeger (SSH) model, physical phase transitions in elastic systems are then calculated. Here, the nearest neighbor (NN) and the next-nearest neighbor (NNN) couplings [50–52] are considered in the 2D SSH model. The coupling strength can be tuned by changing the distance between the lattice points. The intra (inter)-coupling of the unit is controlled by the size of parameter $\beta = (d_2 - d_1)/a$, where $d_1$ and $d_2$ represent the intracellular and intercellular distances, respectively, shown in Fig.1(b). The red dashed lines represent the shrinking-state lattice ($\beta > 0$) and the blue dashed lines represent the expanding-state lattice ($\beta < 0$). In order to characterize physical phase transitions in elastic systems, the elastic real-space quantized quadrupole moment is calculated. The eigen equation of an elastic system in the momentum space is:

$$\mathcal{H}|\psi_n\rangle = E_n|\psi_n\rangle, \qquad (1)$$

where $\mathcal{H}$ is the Hamiltonian of the elastic system, $\psi_n$ is $n$th occupied eigenstate, $E_n$ is the energy eigenvalue. In this 2D SSH model, the Hamiltonian of the elastic system can be expressed as:

$$\mathcal{H} = \mathcal{H}_1 + \mathcal{H}_2, \qquad (2)$$

here

$$\mathcal{H}_1 = \begin{pmatrix} \varepsilon_r & t_{NN1}(\mathbf{k}) & 0 & t_{NN2}(\mathbf{k}) \\ t^*_{NN1}(\mathbf{k}) & \varepsilon_r & t_{NN2}(\mathbf{k}) & 0 \\ 0 & t^*_{NN2}(\mathbf{k}) & \varepsilon_r & t^*_{NN1}(\mathbf{k}) \\ t^*_{NN2}(\mathbf{k}) & 0 & t_{NN1}(\mathbf{k}) & \varepsilon_r \end{pmatrix}, \quad (3)$$

where $t_{NN1}(\mathbf{k}) = w_1 + v_2 e^{-ik_x}$, $t_{NN2}(\mathbf{k}) = w_1 + v_2 e^{-ik_y}$, $t^*_{NN1}(\mathbf{k})$ is the conjugate of $t_{NN1}(\mathbf{k})$, $t^*_{NN2}(\mathbf{k})$ is the conjugate of $t_{NN2}(\mathbf{k})$, $w_1 \propto 1/d_1$ and $v_2 \propto 1/d_2$, $w_1$ and $v_2$ respectively reflects the degree of intracellular and intercellular couplings, $\varepsilon_r$ is the onsite energy of a cylinder on the plate, which can be considered as a resonator. Parameters $w_1$, $v_2$ and $\varepsilon_r$ can be obtained by curve fittings of energy bands from theory and COMSOL (the details are shown in Appendix A), respectively. Besides,

$$\mathcal{H}_2 = \begin{pmatrix} 0 & 0 & t_{NNN1}(\mathbf{k}) & 0 \\ 0 & 0 & 0 & t_{NNN2}(\mathbf{k}) \\ t^*_{NNN1}(\mathbf{k}) & 0 & 0 & 0 \\ 0 & t^*_{NNN2}(\mathbf{k}) & 0 & 0 \end{pmatrix}, \quad (4)$$

where $t_{NNN1}(\mathbf{k}) = w_{n1} + v_{n2} e^{-ik_x} + v_{n4} e^{-ik_y} + v_{n3} e^{-i(k_x+k_y)}$, $t_{NNN2}(\mathbf{k}) = w_{n1} + v_{n2} e^{-ik_x} + v_{n4} e^{-ik_y} + v_{n3} e^{-i(k_x-k_y)}$, $t^*_{NNN1}(\mathbf{k})$ and $t^*_{NNN2}(\mathbf{k})$ are the conjugate of $t_{NNN1}(\mathbf{k})$ and $t_{NNN2}(\mathbf{k})$, respectively, and the parameters $w_{n1}, v_{n2}, v_{n3}$ and $v_{n4}$ reflect the degree of NNN couplings, which can be obtained by curve fittings of energy bands (Appendix A).

To further prove the existence of topological phase transitions in elastic fractal structures, it is necessary to characterize them by using topological invariants. However, the special features of fractal structures make it impossible to characterize those by using chern number [53–56], which is only suitable for periodic structures. In this work, an invariant based on real-space wavefunctions, the elastic real-space quantized quadrupole moment is calculated and used to characterize the topological phase transitions of elastic fractal structures lacking a translation symmetry.

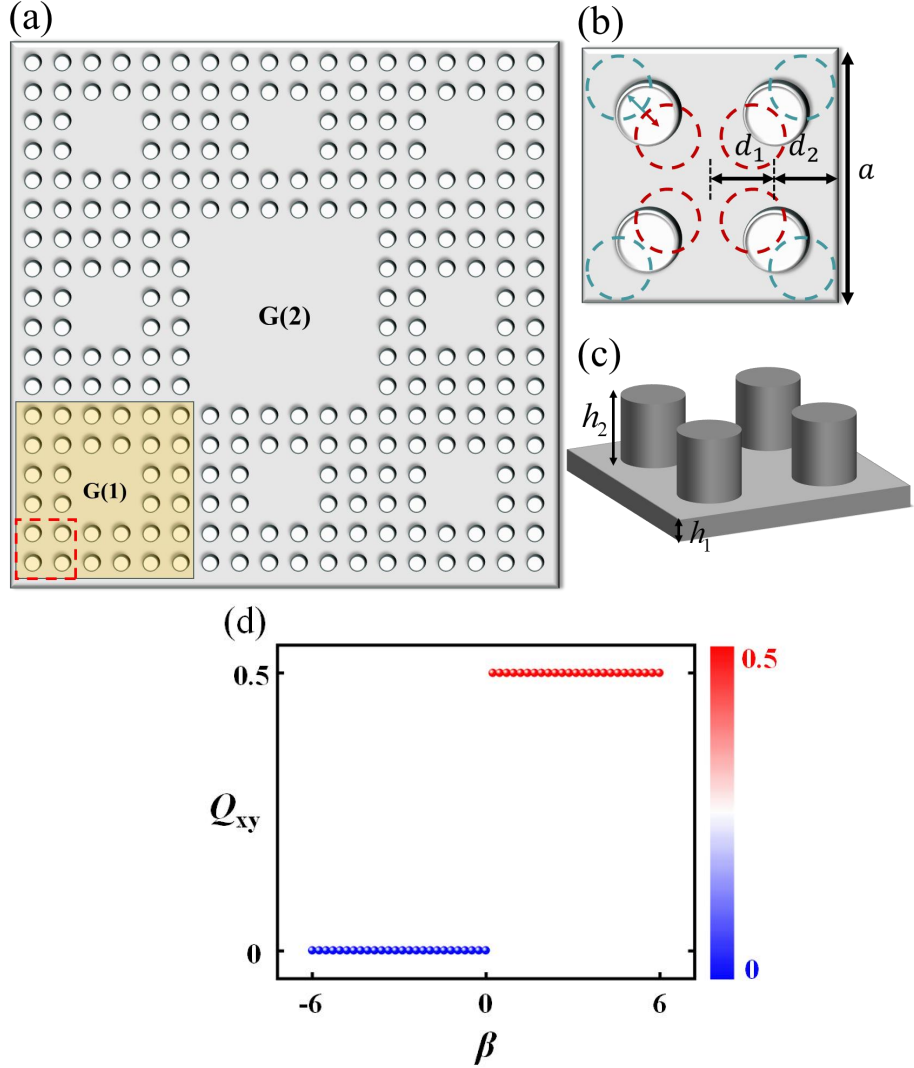

FIG. 1. Elastic fractal model. (a) Schematic of the Sierpinski fractal structure. (b)-(c) Enlarged diagrams of the part inside the red box in Fig.1(a). (d) The elastic real-space quantized quadrupole moment of the Sierpinski fractal structure vs. coupling parameter $\beta$. The topological phase transition is shown by the change of the real-space quantized quadrupole moment, where the value of 0.5 indicates a topological nontrivial system, while the value of 0 represents a trivial system. The system undergoes a topological phase transition at $\beta = 0$.

The quadrupole moment defined in the real space [57] is given by

$$Q_{xy} = \frac{1}{2\pi} \text{Im} \log \left[ \det\left(\Psi_{occ}^{\dagger} \hat{U} \Psi_{occ}\right) \sqrt{\det\left(\hat{U}^{\dagger}\right)} \right] \quad (5)$$

where $\hat{U} = \exp(i2\pi \hat{q}_{xy})$, $\hat{q}_{xy} = \frac{\hat{x}_j \hat{y}_j}{N_x N_y}$, $\hat{x}_j, \hat{y}_j$ are position operators along the $x$ and $y$ directions, respectively, $N_x, N_y$ are the number of units along the $x$ and $y$ directions, respectively. $\Psi_{occ}$ is the wave function of the occupied state in real-space, which can be expressed as:

$$\Psi_{occ} = (|\psi_1\rangle, |\psi_2\rangle, |\psi_3\rangle, \ldots |\psi_{n_c}\rangle). \tag{6}$$

This matrix is a projection operator of the semi-filled occupied state, $\psi_n$ can be calculated from the eigen equation Eq. (1).

For the fractal structure shown in Fig.1(a), the elastic real-space quantized quadrupole moment $Q_{xy}$ is calculated and the result ($Q_{xy}$ vs. $\beta$) is shown in Fig.1(d). The system is topologically protected when $Q_{xy}$=0.5, while it is in a trivial state when $Q_{xy}$=0. It can be seen from Fig.1(d) that the system undergoes a topological phase transition at $\beta = 0$.

## III. Topological states in elastic fractal systems

In this section, the topological states of elastic waves in fractal structures are characterized through a series of simulations and experiments. In this work, only out-of-plane vibrations are considered.

The simulations are conducted via COMSOL Multiphysics, the Solid Mechanics Module is used to model the elastic fractal structures. During FEM simulations, an absorbing boundary condition is applied to four outer edges of the fractal plate to prevent wave reflections from the boundaries.

The fabricated fractal plates are shown in Figs. 2(a) and (b), which are made of aluminum $(\rho = 2700\text{kg/m}^3, v = 0.33, E = 68.9\text{GPa})$. The exciter is located at the position of the yellow pentagram shown in Figs. 2(a) and (b). The vibration absorbing glue is applied around the sample to ensure that the energy is not excessively dissipated.

The experimental platform is shown in Fig. 2(c). Periodic sinusoidal wave signals with frequencies from 85-135 kHz are generated by using a signal generator (DG1022U, RIGOL), which is then amplified by a power amplifier (2706, B&K) to drive the piezoelectric exciter (a PZT resonator with a radius of 5 mm). The scanning laser vibrometer (PSV-500, POLYTEC) is used to measure the out-of-plane velocities at selected points throughout the region, which are then transmitted to the vibrometer controller and converted into displacement signals.

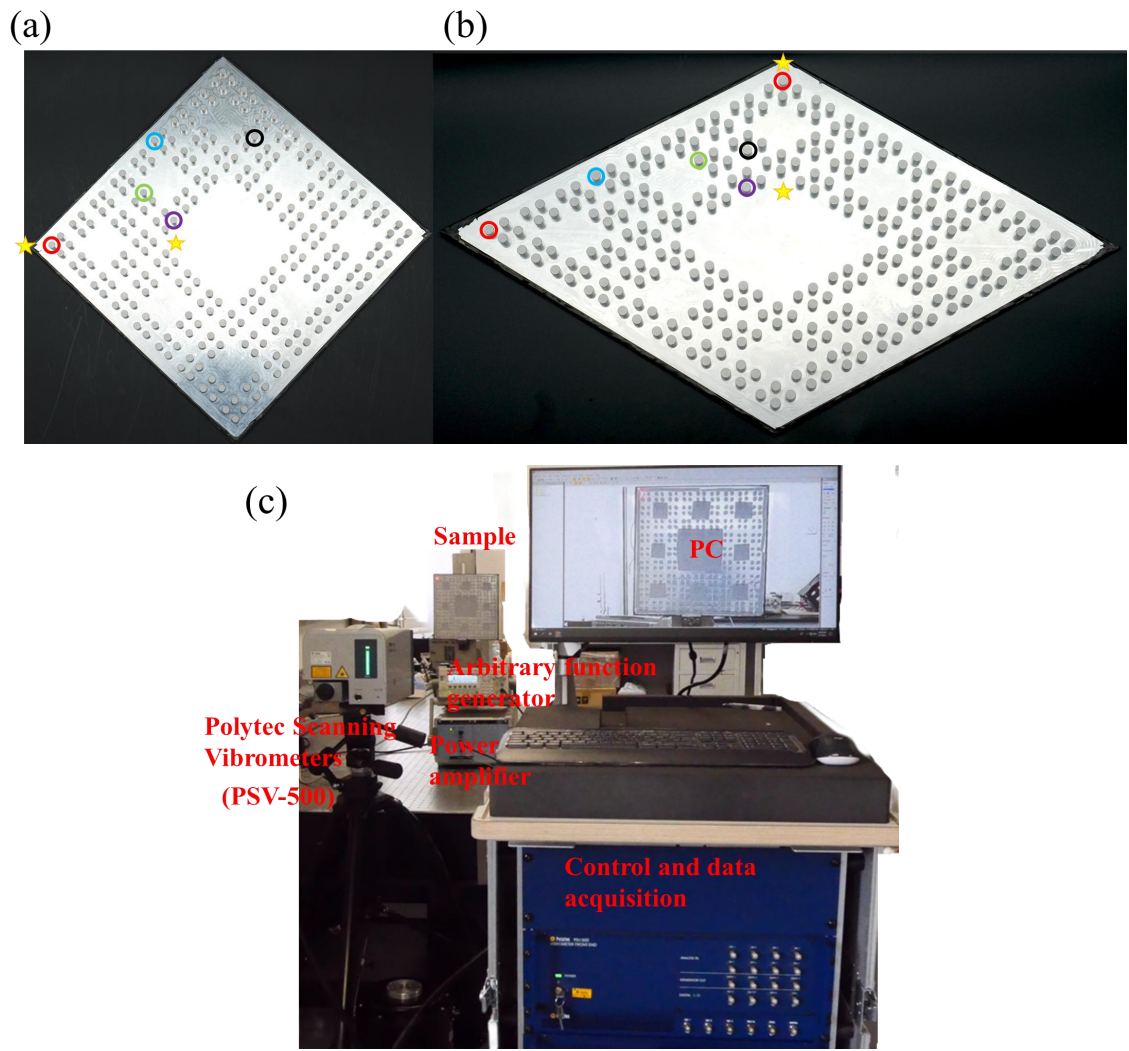

FIG. 2. Elastic fractal plates and the experimental platform. (a) and (b) Fabricated Sierpinski and rhombus elastic fractal plates, respectively. The yellow pentagram represents the position of the piezoelectric exciter, and red, blue, purple, green and black points indicate the frequency measuring positions of the outer corner state, outer edge state, inner edge state, inner corner state and bulk state, respectively. (c) The experimental platform.

## A. Topological states of the Sierpinski fractal structure

For the G(2) Sierpinski fractal structure shown in Fig.1, when $\beta = 2$ ($w_1 = 0.35$), $Q_{xy} = 0.5$, the elastic system is in a non-trivial state. The simulation and experimental results of eigenstates of the system are shown in Fig.3. In Fig.3(a), it can be seen that there exist 100 bulk states (black), 72 inner corner states (green), 16 inner edge states (purple), 64 outer edge states (blue) and four outer corner states (red). Fig.3(c)-(g) shows displacement distributions of the bulk states (82.217 kHz), inner edge states (112.63kHz), inner corner states (97.153kHz), and outer edge states (119.63kHz), outer corner states (126.00kHz) obtained by numerical simulations, respectively. It is shown that for the inner edge state, inner corner state, outer edge state and outer corner state, the energy is highly concentrated in the inner side, inner corner, outer side and outer corner regions, respectively. For the bulk states, the energy is distributed all over the plane.

By using the box-counting method, dimension numbers of the Sierpinski fractal structure have been calculated, and based on that, the numbers of various states have been also calculated, the results are shown in Appendix B. It is shown that the numbers of various states obtained by simulations is consistent with that calculated based on the dimension.

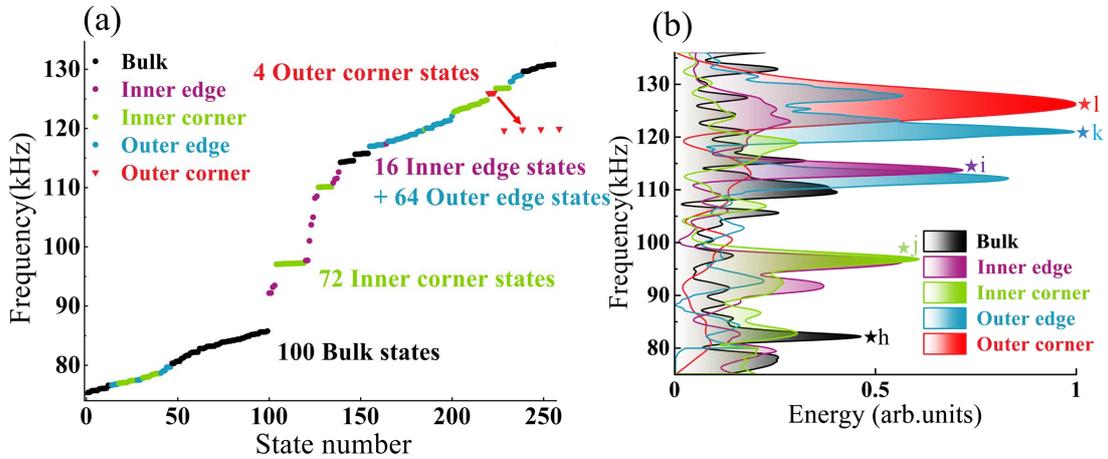

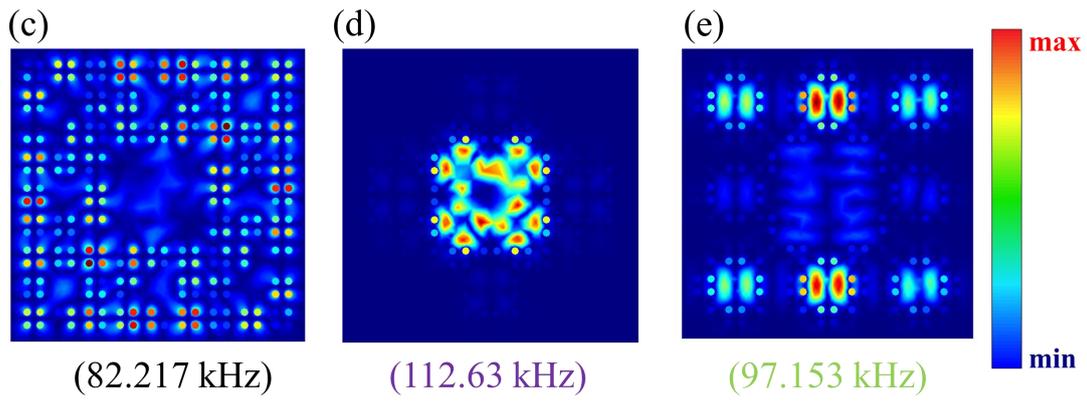

(c) (82.217 kHz)  (d) (112.63 kHz)  (e) (97.153 kHz)

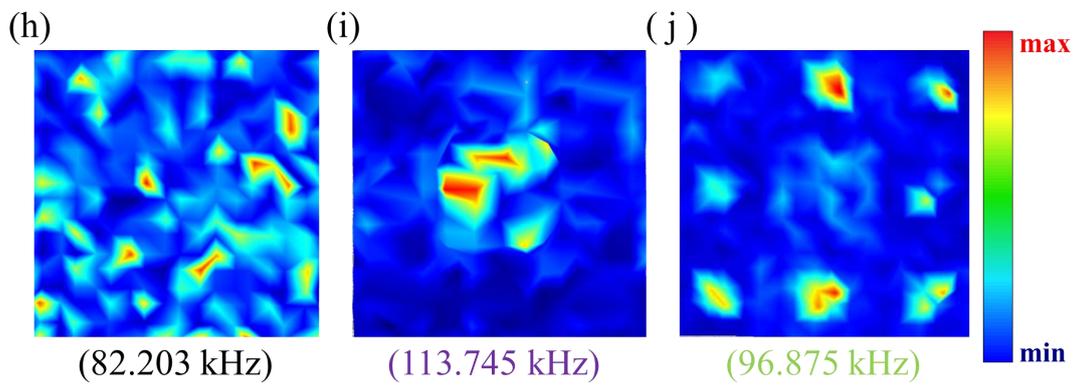

(h) (82.203 kHz)  (i) (113.745 kHz)  (j) (96.875 kHz)

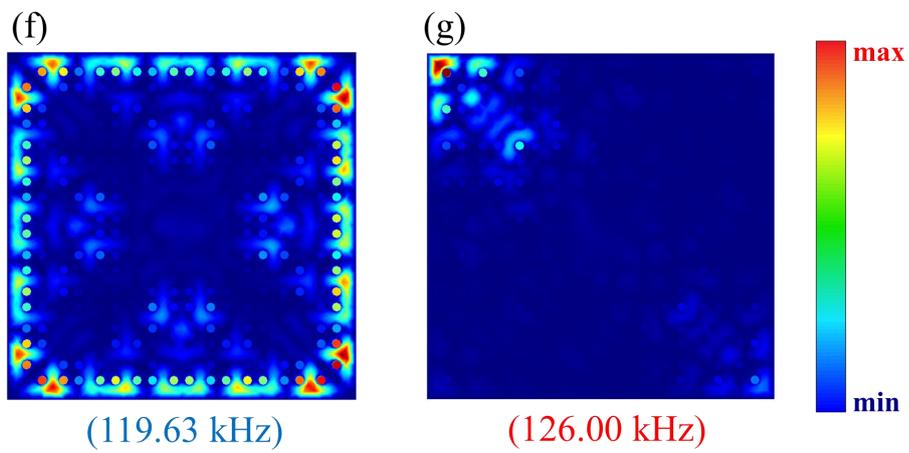

(f) (119.63 kHz)  (g) (126.00 kHz)

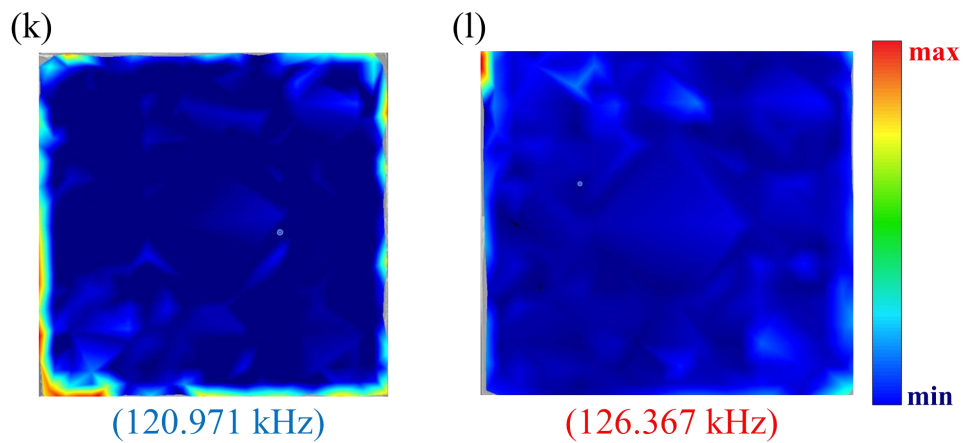

(k) (120.971 kHz)  (l) (126.367 kHz)

FIG. 3. Results from experiments and numerical simulations of topological eigenstates of the G (2) Sierpinski fractal structure at non-trivial state. (a) The eigen frequencies obtained for $\beta=2$ by simulations. (b) Normalized spectral diagram obtained by experiments. (c)-(g) Simulated displacement distributions of the bulk state at 82.217kHz, the inner edge state at 112.63 kHz, the inner corner state at 97.153 kHz, the edge state at 119.63 kHz and the topological corner state at 126 kHz. (h)-(l) The experimental results of displacement distributions of the bulk state at 82.203 kHz, the inner edge state at 113.745 kHz, the inner corner state at 96.875 kHz, the outer edge state at 120.971 kHz and the outer corner state at 126.367 kHz.

Elastic fractal plates are fabricated and the according tests are carried out to verify the simulation results. The material and structural parameters in experiments are the same as those in simulations. The displacement response spectrums are normalized according to the maximum value, which are shown in Fig.3(b). It can be seen that for the corner and edge states, and the frequencies are close to the those obtained by numerical simulations in Fig.3(a). Fig.3(h)-(l) shows displacement distributions of bulk state (82.203 kHz), inner edge state (113.745 kHz), inner corner state (96.875kHz), and outer edge state (120.971 kHz) and outer corner state (126.367 kHz) obtained by experiments. It is shown that displacement distributions of various states obtained by experiments are in approximate agreement with the simulation results.

Especially, different from the acoustic fractal system, for the elastic fractal system, the outer and inner edge states can emerge separately. For the acoustic fractal system, out of the inner and outer boundaries, the mediums are both air, the coupling boundary conditions are the same, thus the eigen frequencies of the inner and outer edge states are coincident, namely the outer and inner edge states can not emerge separately. While, for the elastic fractal system, the outside of the outer boundary is air, there is still a transmission medium for elastic waves inside the inner boundary of the elastic fractal plate (it is aluminum plate in this work), namely the coupling boundary conditions are different, thus the eigen frequencies corresponding to the inner and outer edge states do not coincide, namely the outer and inner edge states can emerge separately by tuning the operating frequency. This characteristic is vital for elastic-wave devices used in the integrated sensing and particle manipulation in microfluidics to realize flexible sensing and control functions.

Due to the particularity of the fractal structure, the eigenstates of non-trivial elastic fractal systems ($\beta > 0$, namely $w_1 > 0.1$) contain topological states, while the trivial elastic fractal systems ($\beta < 0$, namely $w_1 < 0.1$) contain the defect states. To distinguish the two kinds of states, the eigenstates of trivial systems are shown in Fig.C1 in Appendix C. The results show that the inner edge and inner corner states occur in both topological and trivial systems. By checking the robustness of these states, non-topologies of those in trivial systems are confirmed in Appendix C, and topologies of those in topological systems are confirmed in Appendix D. The topologies of outer corner states in topological systems are analyzed in sub-Section C.

**B. Topological states of the rhombus fractal structure**

To check the influences of the geometry of the fractal structure on the robustness of high-order topological states in sub-Section C, the topological states of the elastic rhombus fractal structure are analyzed in this sub-Section.

The elastic rhombus fractal model is shown in Fig.C3 of Appendix C, and the eigenstates of the G(2) rhombus fractal structure are numerically simulated, the results are shown in Fig.4. When the intracellular coupling is larger than the extracellular coupling, the system enters the topological state. Fig. 4(a) shows eigen frequencies obtained by simulations, bulk states, inner edge states, inner corner states, $2\pi/3$ outer corner states, $\pi/3$ outer corner states, and outer edge states are marked with black dots, purple dots, green dots, red dots, pink dots and blue dots, respectively. Fig.5(c)-(h) show displacement distributions of bulk state (85.22 kHz), inner edge state (95.819 kHz), inner corner state (113.28 kHz), outer edge state (126.81 kHz), $\pi/3$ outer corner state (98.054 kHz) and $2\pi/3$ outer corner state (115.83 kHz) obtained by numerical simulations. It is shown that for the inner edge, inner corner, outer edge and outer corner states, the energy distribution characteristics are obvious.

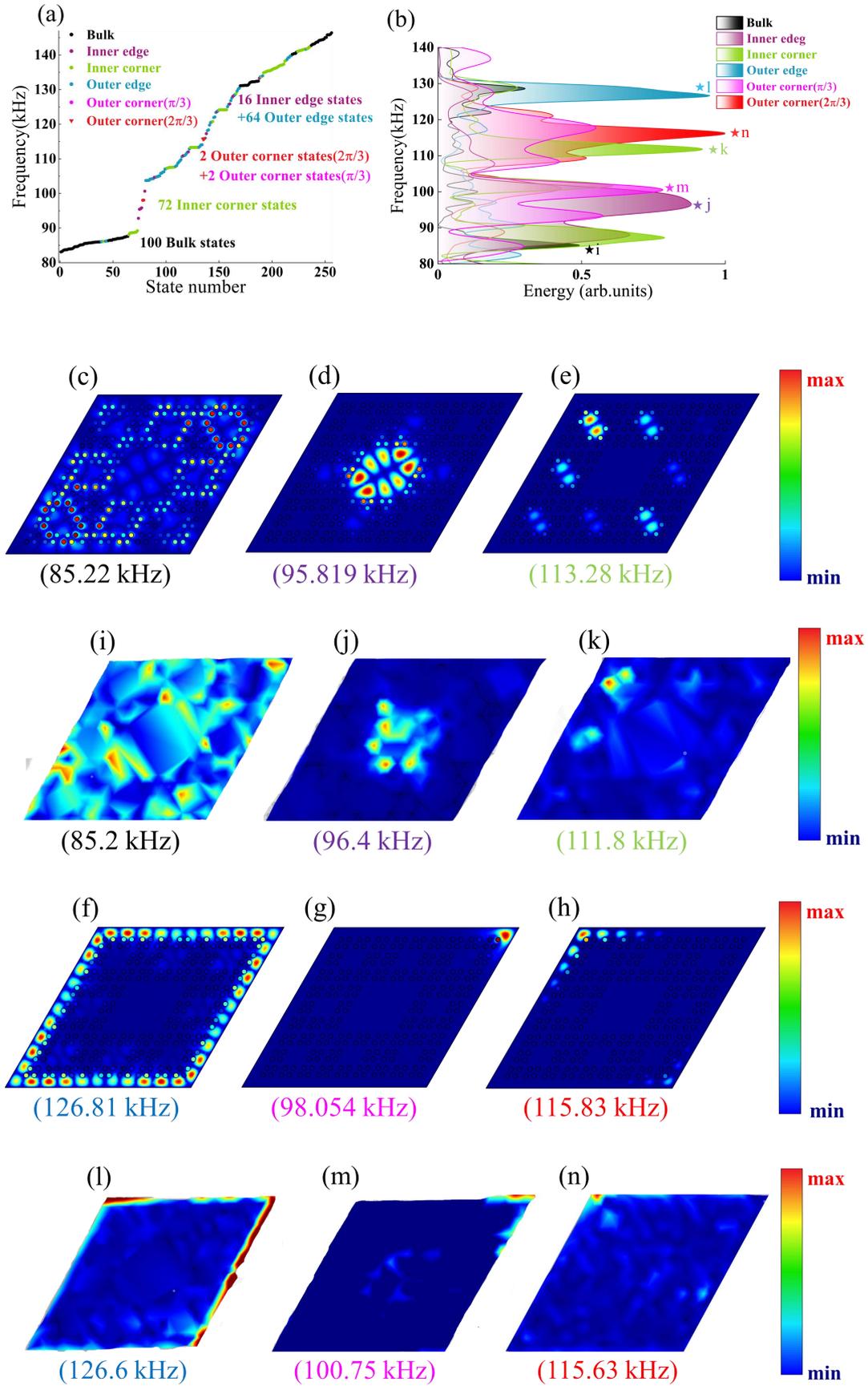

FIG. 4. Results from experiments and numerical simulations of topological eigenstates of the G (2)

rhombus fractal structure at non-trivial state. (a) The non-trivial eigen frequencies obtained by simulations. (b) The measured normalized spectral diagram. (c)-(g) Simulation displacement distributions of the bulk state at 85.22 kHz, the inner edge state at 95.819 kHz, the inner corner state at 113.28 kHz, the edge state at 126.81 kHz, the $\pi/3$ outer corner state at 98.054 kHz and the $2\pi/3$ outer corner state at 115.83 kHz. (h)-(l) The experimental results of displacement distributions of the bulk state at 85.2 kHz, the inner edge state at 96.4 kHz, the inner corner state at 111.8 kHz, the edge state at 126.6 kHz, $\pi/3$ outer corner state at 100.75 kHz and the $2\pi/3$ outer corner state at 115.63 kHz.

The fabricated rhombus fractal samples are shown in Fig. 2(b). The measured normalized response spectrum curves are shown in Fig.4(b). It can be seen that the maximum energy appears at the outer corner states, and the corresponding frequencies of edge and corner states are consistent with those obtained by numerical calculations.

To distinguish topological and trivial modes of the rhombus structure, the eigenstates of the trivial system are shown in Fig. C4(a) in Appendix C. It is shown that the inner edge and inner corner states occur in both topological and trivial systems. Non-topologies of those in trivial systems are confirmed in Appendix C.2 by checking the robustness of these states, and topologies of those in topological systems are confirmed in Appendix D. The topologies of outer corner states in topological systems are analyzed in sub-Section C.

### C. Robustness verification and comparison of topological corner states in Sierpinski and rhombus elastic fractal structures

The topologies are verified by detecting the robustness of topological corner states to structure defects and lattice-disorders. For the G(2) Sierpinski fractal structure, the columns at the diagonal points are lengthened to introduce disorders, as shown in the inset in Fig. 5(c); Besides, disorders are also introduced by moving two columns to change the coupling strength (coupling parameter $\beta$ is changed to 4), as shown in the inset of Fig.5(d); For the defect setting, three columns near the corner points are deleted, forming a cavity, as shown in the inset in Fig.5(e). Fig.5(a) shows the normalized energy spectrum curves of corner states obtained by experiments for the Sierpinski fractal plate with disorder (column-elongation), disorder (column-moving), defect (column-deleting) and no defect or disorder, respectively. The results show that after the introductions of defects and disorders in the G(2)

Sierpinski fractal structure, although there is a small fluctuation for the frequencies of corner states, and some energy spread to the edge, the energy is still concentrated near the corner region, indicating that the introduction of defects and disorders have slight influence on the energy distributions of corner states, thus the robustness of the topological corner states are obvious.

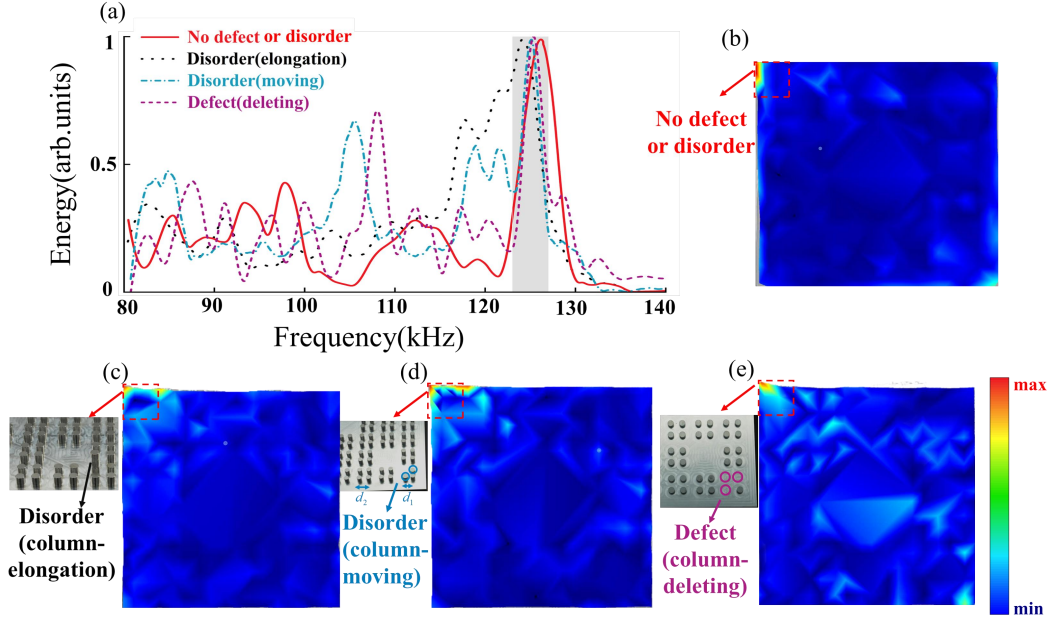

FIG. 5. Robustness verification of the topological corner states of the G(2) Sierpinski fractal structure. (a) Normalized energy spectra. (b)-(e) The measured displacement distributions of Sierpinski fractal plates with no defect and disorder, with disorder (column-elongation), with disorder (column-moving), with defect (column-deleting), respectively.

For the rhombus fractal structure, two outer corner states are at the $2\pi/3$ corner, and the other two are at the $\pi/3$ corner. In analogy to electromagnetic waves, the topological index N [58, 59] is calculated to verify the topology of the corner states. The details can be found in Appendix G. The results show that for the $\pi/3$ corner, the topological index N=0, implying the existence of trivial corner state. While for the $2\pi/3$ corner, the topological index N=1, implying the existence of topologically protected corner states.

For the outer corner states at the $2\pi/3$ corner, defects and disorders are introduced in the same way as those for the Sierpinski fractal structure, the measured displacement distributions and normalized energy spectra are shown in Fig.6. It is shown that after the disorders and defects are introduced, the energy concentration effects at the $2\pi/3$ corner are basically unchanged. Thus, the corner states at $2\pi/3$ corner of the rhombus elastic fractal plate are topological, and strongly robust to defects and disorders.

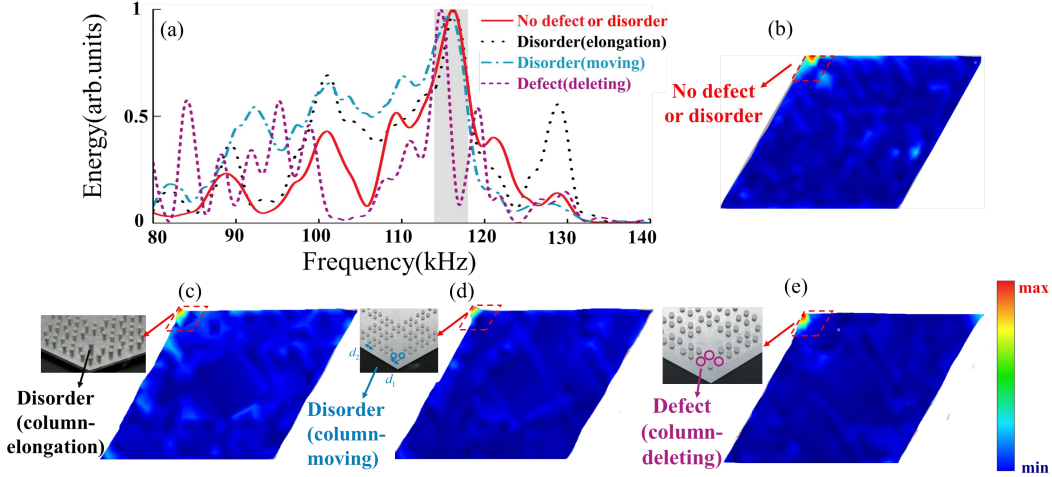

FIG. 6. Robustness verification of $2\pi/3$ topological corner states of the G(2) rhombus fractal structure. (a) Normalized energy spectra. (b)-(e) The measured displacement distributions of rhombus fractal plates with no defect and disorder, with disorder (column-elongation), with disorder (column-moving), with defect (column-deleting).

The above results show that, for the rhombus fractal structure, the influences of the defects and disorders on the energy concentration degree of topological corner states are less obviously than those for the Sierpinski fractal structure shown in Fig.5. Therefore, the robustnesses of topological corner states of the rhombus fractal structure are significantly better than that of the Sierpinski fractal structure. In reference [60], it was shown that there is indeed an obvious difference in quantified corner charge due to the different corner shapes and symmetries. In our rhomboid fractal model, at $\pi/3$ corners and $2\pi/3$ corners, the difference in corner charges is caused by the difference in coupling connections at two kind of corners, which affects the uneven distribution of corner charges at four outer corners, so that the topology of $\pi/3$ corner states disappears and that of $2\pi/3$ corner states increases.

In order to prove that the corner state at the $\pi/3$ corner of the rhombus structure is trivial, experimental robustness verifications are conducted for the $\pi/3$ corner of the rhombus fractal plate. The defects and disorders are introduced to observe whether the robustness exists, which are shown in the inset of Fig.7. Displacement distributions are shown in Fig.7. It is found that after defects and disorders are introduced, the corner state appearing at the $\pi/3$ corner does not maintain the energy concentration, and the energy spreads around obviously, namely there is no robustness. Thus, the corner state appearing at the $\pi/3$ corner of the rhombus fractal structure is indeed a trivial corner state.

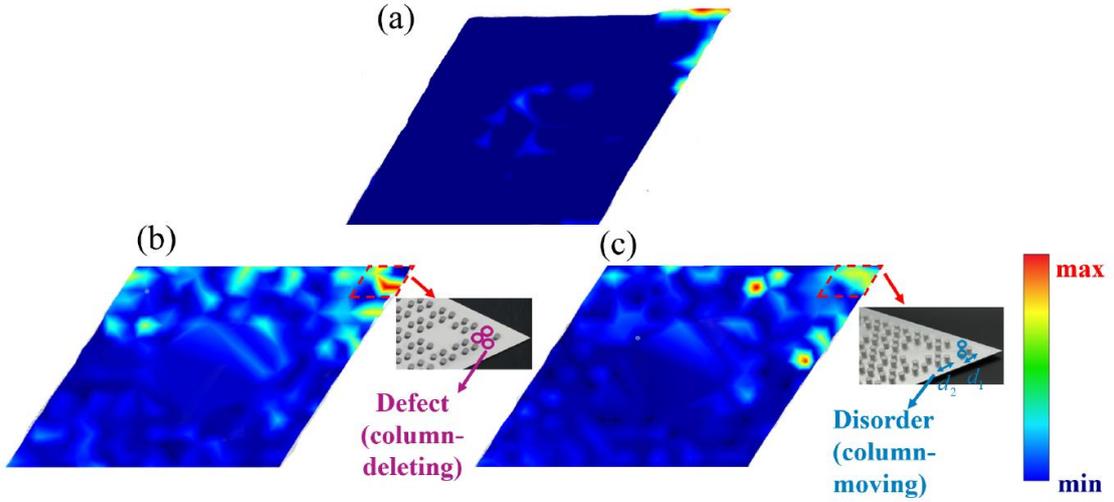

FIG. 7. Experimental verification of the absence of topological protection of $\pi/3$ corner states of the G(2) rhombus fractal structure. (a)-(c) Displacement distribution of $\pi/3$ corner states: (a) With no defect or disorder; (b) With defect (column-deleting); (c) With disorder (column-moving).

In addition, eigenstates calculations are also carried out for the model of G(3) states, and the results are shown in Appendix E.

### D. Comparison of richness of topological states in fractal and periodic structures

In order to verify further the superiority of fractal structures in elastic topological system, square periodic structures with an integer dimension are designed (Fig. F1(a) in Appendix F), and the coupling parameters of the structure are consistent with the fractal structure. The details of simulation results are shown in Appendix F.

It is found that due to the disappearance of the self-similarity of the structure, the periodic structure has far fewer topological states (only 28) than the Sierpinski fractal structure (156). Rich topological states are the key to realize the application of topological elastic-wave devices in the field of integrated sensing, particle manipulation, energy harvesters and high Q-factor resonators, and can significantly expand the operating frequency range and improve the adaptation of the devices to various loadings.

Moreover, a large number of topological inner corner states appear in the elastic fractal structure, which provides a good method for realizing the elaborate and precise energy aggregation of elastic waves, and can be applied to control particle aggregation in microfluidics.

## IV. Conclusions

In this work, higher-order topological states in elastic fractal systems are investigated. The elastic real-space quantized quadrupole moment is calculated and used to characterize the topological phase transitions of elastic fractal structures, and formation conditions of topological phase transitions in elastic fractal systems are revealed; The topological corner and edge states of elastic waves in fractal structures are realized numerically and experimentally. The results show that the richness of topological states in elastic fractal structures is much higher than that of topological insulators based on periodic structures, which is vital in integrated sensing, particle manipulation and energy acquisition applications. The robustness of the topological states in elastic fractal structures are verified by introducing structure defects and lattice disorders. The results show that the robustness of topological corner states in rhombus fractal structures is obviously stronger than that in Sierpinski fractal structures, and the physical mechanism is clarified.

The topological phenomena in elastic fractal systems revealed in this work brings new insights to topological physics of elastic wave-motion systems, and breaks through the limitation of topological states relying on periodic elastic structures. Topological phenomena based on fractal structures can be applied in high-resolution elastic-wave energy locations, high-sensitivity detections, high-Q resonating, elastic energy acquisition, and other related fields, which have important application prospects and can bring new research ideas to the above fields.

**Declaration of competing interest**

The authors declare that they have no known competing financial interests or personal relationships that could have appeared to influence the work reported in this paper.

**Data availability**

Data will be made available on request.

**Acknowledgments**

This work was supported by the National Natural Science Foundation of China (Nos. 12172183。12302113), National Key Research and Development Program of China (No. 2023YFE0111000), the Natural Science Foundation of Zhejiang Province (No. LZ24A020001), International Science and Technology Cooperation Project launched by Science and Technology Bureau of Ningbo City, Zhejiang Province, China (No. 2023H011), Ningbo Municipal Natural Science Foundation (No. 2022J090)，One health Interdisciplinary Research Project (No. HY202206), Ningbo University.

# Appendix A. The Hamiltonian of SSH model and parameter fittings

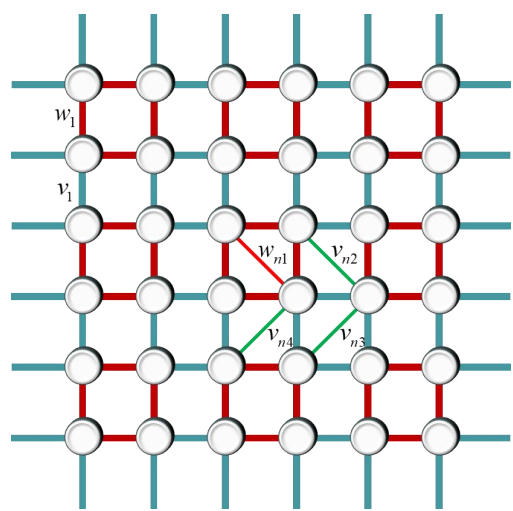

FIG. A1. The square SSH model with NN ( $w_1$ and $v_1$ ) NNN couplings ( $w_{n1}, v_{n2}, v_{n3}$ and $v_{n4}$ ). Four lattice sites in the unit cell form a symmetric quadrate. Red lines

represent intracellular couplings and green lines represent intercellular couplings.

The energy bands are calculated based on the Hamiltonian expression Eqs.(3) and (4), and those are also calculated by COMSOL, based on which, the curve fittings are carried out to obtain the coupled parameters of the Hamiltonian $W_1, V_1, W_{n1}, V_{n2}, V_{n3}$ and $V_{n4}$. Here, Genetic Algorithm (GA) are used to obtain the best fitting value through fast iterative calculation, and the fitting value reaches the best after 1000 iterations. The values of coupling parameters in the Hamiltonian, and the fitted energy band curves are shown in Fig.A2 (the red curves are the theoretical results and the blue ones are the results from COMSOL).

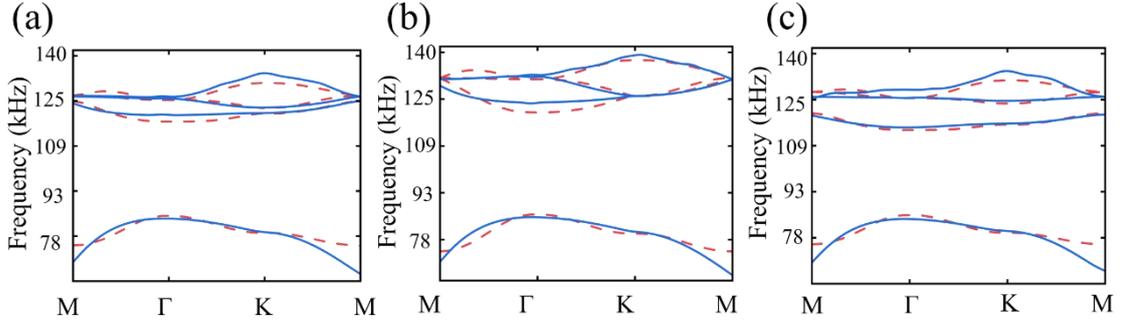

FIG. A2. Fitted energy band curves and the coupled parameters of the Hamiltonian. the red curves are the theoretical results and the blue ones are the results from COMSOL. (a) $\beta = -0.4, w_1 = -0.53, v_2 = 0.111, w_{n1} = 0.733, v_{n2} = v_{n4} = v_{n3} = -0.02$ and $\varepsilon_r = 5.806$.

(b) $\beta = 0, w_1 = -0.587, v_2 = 0.153, w_{n1} = 0.737, v_{n2} = v_{n4} = v_{n3} = -0.005$ and $\varepsilon_r = 6.021$. (c) $\beta = 0.4, w_1 = 0.47, v_2 = -0.1, w_{n1} = 0.728, v_{n2} = v_{n4} = v_{n3} = -0.027$ and $\varepsilon_r = 5.78$.

## Appendix B. The dimensions of the fractal structures and the number of the topological states

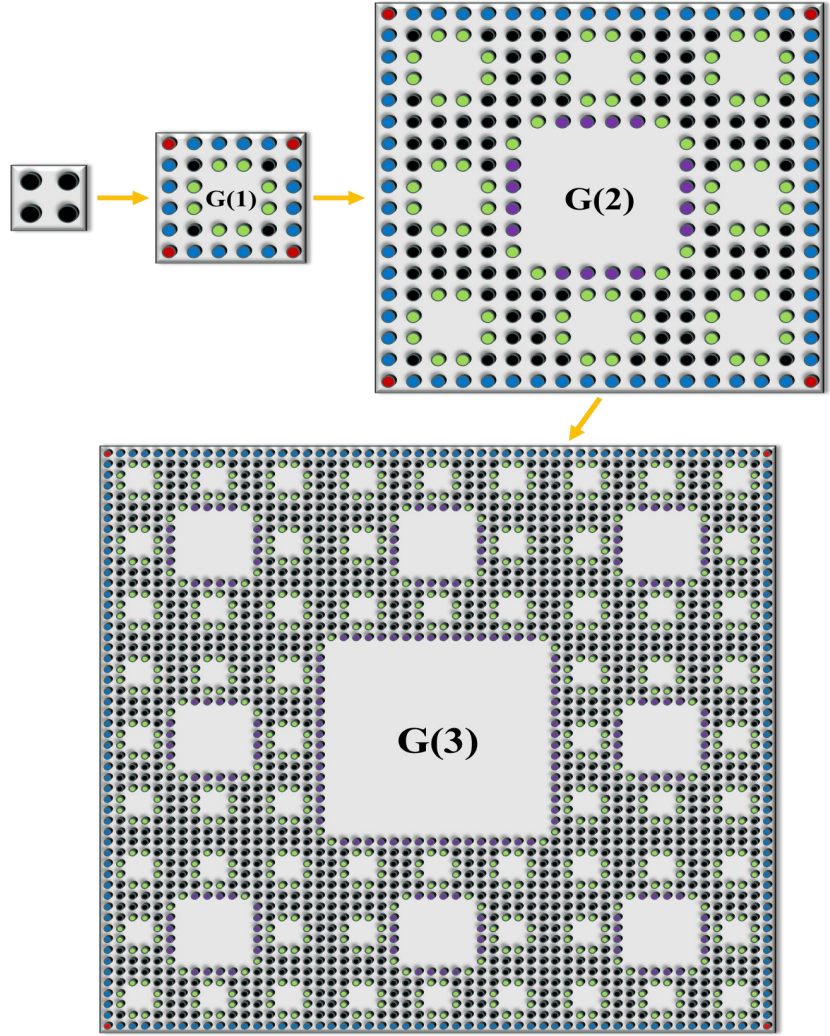

FIG. B1. Schematic diagrams of the fractal structure. The smallest cell in the fractal structure has four grid points. The first generation G(1) has eight minimal cells, the second generation G(2) contains eight G(1)s, and G(3) contains eight G(2)s, or 64 G(1)s. The red, blue, purple, green, and black columns in the figure denote the outer corner state, the outer edge state, the inner edge state, the inner corner state, and the bulk state, respectively.

The box-counting method [49] is used to calculate the dimension number of the fractal structures. As can be seen from Fig.B1, the total number of boxes in each generation is $N_{G_n} = 4 \times 8^n$, and the number of transverse boxes is $N_L = 2 \times 3^n$, so the fractal dimension can be obtained:

$$\dim(D_G) = \lim_{n \to \infty} \frac{\ln(N_{G_n})}{\ln(N_L)} = \lim_{n \to \infty} \frac{\ln(4 \times 8^n)}{\ln(2 \times 3^n)} \approx 1.893, \quad \text{(B.1)}$$

$$\dim(D_o) = \lim_{n\to\infty} \frac{\ln(N_c)}{\ln(N_L)} = \lim_{n\to\infty} \frac{\ln(4)}{\ln(2\times 3^n)} = 0, \qquad (B.2)$$

$$\dim(D_i) = \lim_{n\to\infty} \frac{\ln(N_i)}{\ln(N_L)} = \lim_{n\to\infty} \frac{\ln(\sum_{i=1}^{n} 8^i)}{\ln(2\times 3^n)} \approx 1.893, \qquad (B.3)$$

$$\dim(D_e) = \lim_{n\to\infty} \frac{\ln(N_e)}{\ln(N_L)} = \lim_{n\to\infty} \frac{\ln(8^n \times (3^n-1) + \sum_{i=1}^{n-1}(3^i-1)\times 8^{n-i})}{\ln(2\times 3^n)} \approx 1.893, \qquad (B.4)$$

where $N_c, N_i, N_e$ are numbers of outer corner, inner corner and edge states, respectively. For the fractal structure, the total number of states of G(1) fractal structure are 32, including 4 outer corner states, 8 inner corner states, and 16 edge states. For G(2), the total box number is consistent with the total lattice number, which is 256. In the above calculation, it can be seen that there are 4 outer corner states, 72 inner corner states and 80 edge states, which is consistent with the calculation results in the Section 3.

Extending the fractal structure to an infinite space, the proportions of the numbers of inner corner, edge, and bulk states to the total number of states generated in infinite iterations are:

$$\omega_{inner\ corner} = \lim_{n\to\infty} \frac{N_{ic}}{N_{G_n}} = \lim_{n\to\infty} \frac{\sum_{i=1}^{n} 8^i}{4\times 8^n} = \frac{2}{7} \qquad (B.5)$$

$$\omega_{edge} = \lim_{n\to\infty} \frac{N_e}{N_{G_n}} = \lim_{n\to\infty} \frac{8^n \times (3^n-1) + \sum_{i=1}^{n-1}(3^i-1)\times 8^{n-i}}{4\times 8^n} = \frac{4}{35} \qquad (B.6)$$

$$\omega_{bulk} = 1 - \omega_{inner\ corner} - \omega_{edge} = \frac{3}{5} \qquad (B.7)$$

# Appendix C. Trivial states of fractal structures

## C.1 Trivial states of the Sierpinski fractal structure

As shown in Fig.1(e), when ($\beta < 0$, namely $w_1 < 0.1$), $Q_{xy} = 0$, the system is under a trivial state. The simulation results of the eigenstates of the Sierpinski fractal structure for $\beta = -2 (w_1 = -0.45)$ are shown in Fig.C1(a). Inner edge states, inner corner states and bulk states are represented by purple, green, and black dots, respectively. These inner edge and corner states are caused by the absence of lattice points in the structure, rather than a nontrivial state of the system. Displacement distributions of bulk states (115.91kHz), inner edge states (113.09 kHz) and inner corner states (104.03 kHz) are respectively shown in Fig.C1(b), (c) and (d), respectively. For the bulk state, the energy is distributed throughout the structure and does not show a concentration state; while for the inner corner state and inner edge state, the energy gathers at the missing grid points of the fractal structure.

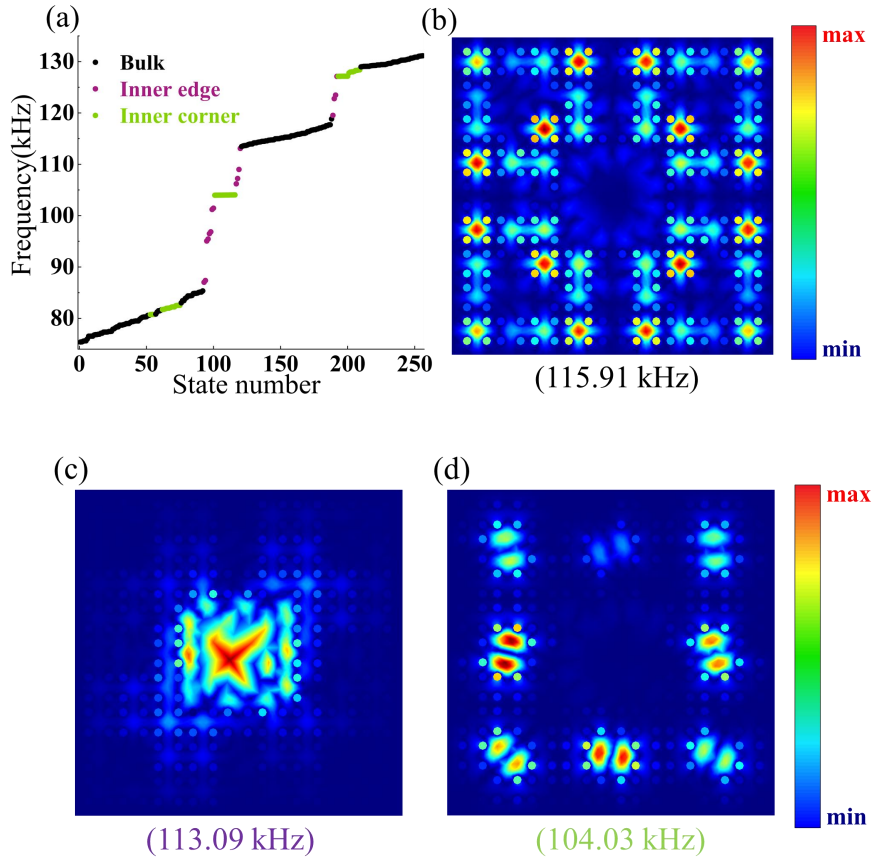

FIG. C1. Trivial states of the Sierpinski fractal structure. (a) The eigen frequencies of the G(2) Sierpinski fractal structure with a coupling parameter $\beta = -2$ obtained by simulations. Displacement distribution of eigenstates: (b) the bulk state at 115.91 kHz, (c) the inner edge states at 113.09 kHz, and (d) the inner corner states at 104.03 kHz.

For the G(2) Sierpinski fractal structure at trivial state, disorders are set to verify the non-robustness of the trivial states. Disorders are introduced by add columns to change the coupling strength. For the inner edge state, two cylinders are added at each corner of the inner boundary, as shown in Fig.C2(b). For the inner corner state, add two columns to the diagonal corners respectively at the inner corner, as shown in the inset of Fig.C2(e). Fig.C2(c) and (f) show the displacement distributions of inner edge and corner states with disorders obtained by simulations. It can be seen that after the introductions of disorders in the G(2) Sierpinski fractal structure, the energy is no longer concentrated near the inner edges and corners. The results indicate that the inner edge and corner states of the G(2) Sierpinski fractal structure at trivial state have no robustness.

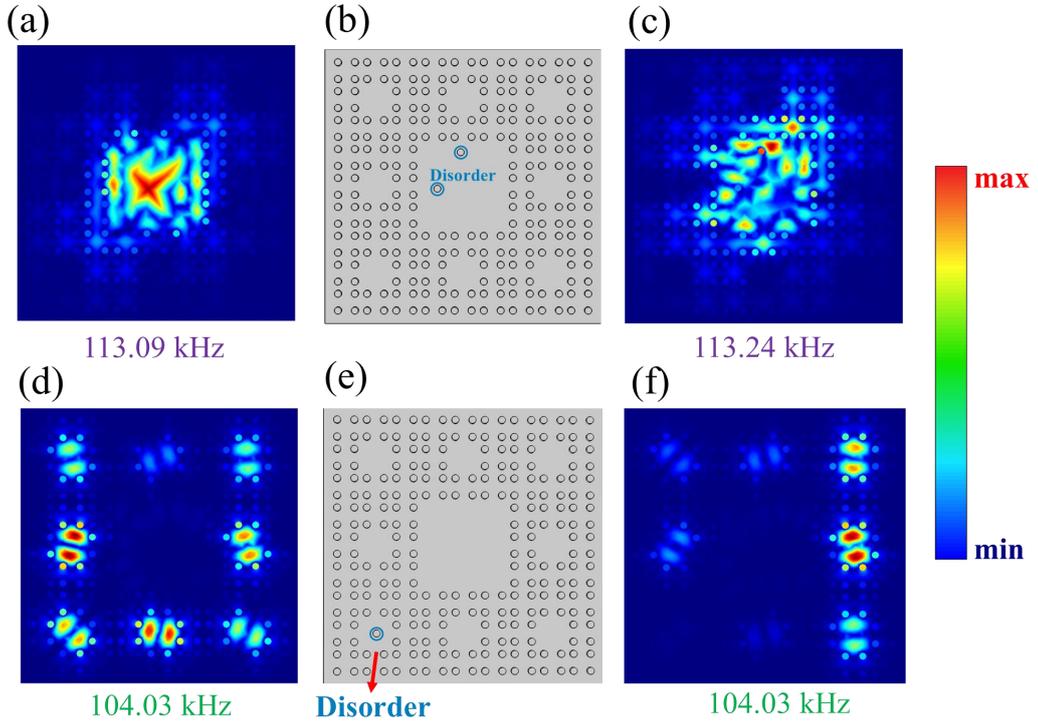

FIG. C2. Verification of the non-robustness of trivial edge and corner states of the Sierpinski fractal structure. (a) Edge state with no lattice disorder. (b) Setting of disorders for edge state. (c) Edge state with lattice disorders. (d) Corner state with no lattice disorder. (e) Setting of disorders for corner states. (f) Corner state with lattice disorders.

## C.2 Trivial states of the rhombus fractal structure

The elastic rhombus fractal model is shown in Fig. C3. Eigen frequencies of the G(2) rhomboid fractal structure obtained by simulations are shown in Fig. C4(a). When the intracellular coupling is smaller than the intercellular coupling (expansion state), the system is under the trivial state. The purple, green and black dots represent the inner edge states, inner corner states, and bulk states, respectively. The appeared inner corner states are caused by the characteristics of the structure itself, rather than a nontrivial state of the system. Displacement distributions of bulk states, inner states and corner states are shown in Fig. C4(b), (c) and (d), respectively. For bulk, inner edge and corner states, the according displacement distribution characteristics are obvious.

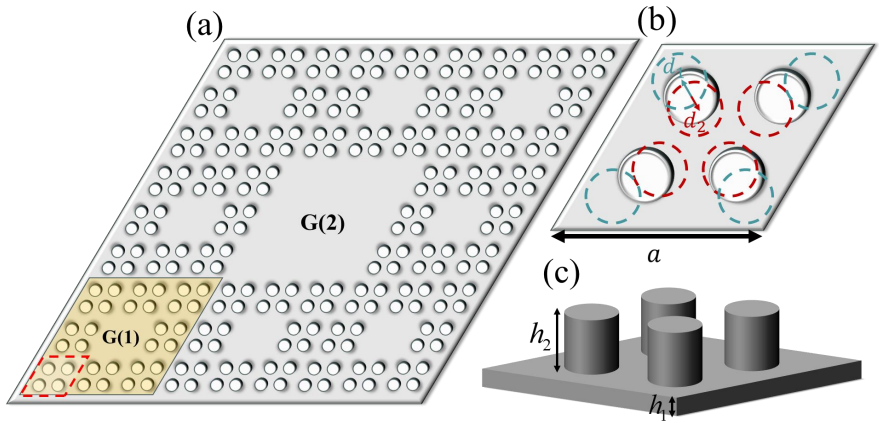

FIG. C3. Elastic rhombus fractal model. (a) Schematic of the rhombus fractal structure. (b)-(c) Enlarged diagram of the part inside the red box in (a).

For the rhombus fractal structure, lattice disorders are introduced in the same way as those for the Sierpinski fractal structure, the displacement distributions are shown in Fig. C5. Fig. C5(c) and (f) show the displacement distributions of inner edge and corner states with disorders obtained by simulations. It can be seen that after the introductions of disorders, the energy is no longer concentrated near the inner edge and corners. The results indicate that the trivial states have no robustness.

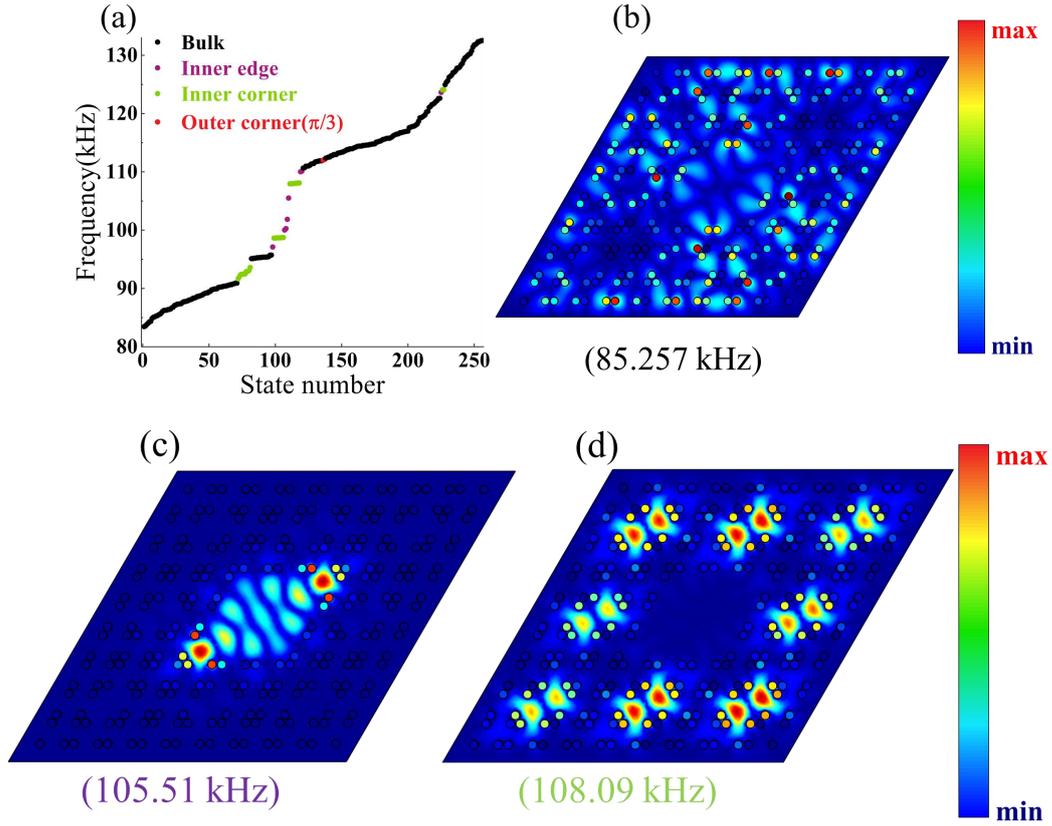

FIG. C4. Trivial system of the rhombus fractal structure. (a) The eigen frequencies of the G (2) rhombus fractal structure at the trivial state obtained by simulations. Displacement distribution of eigenstates: (b) The bulk state at 85.257 kHz. (c) The inner edge state at 105.51 kHz. (d) The inner corner state at 108.09 kHz.

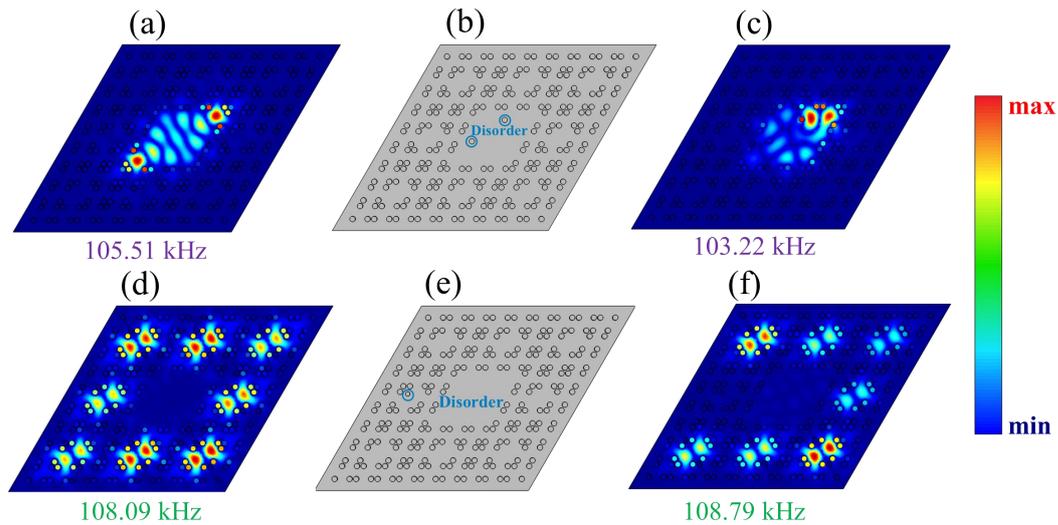

FIG. C5. Verification of the non-robustness of trivial edge and corner states of the rhombus fractal structure. (a) Edge state with no lattice disorder. (b) Setting of disorders for edge state. (c) Edge state with lattice disorders. (d) Corner state with no lattice disorder. (e) Setting of disorders for corner states. (f) Corner state with lattice disorders.

# Appendix D. Robustness of the topological states of the elastic fractal structures

For G(2) Sierpinski and rhombic elastic fractal structures, defects and disorders are set to verify the robustness of the topological inner corner, inner edge and outer edge states.

Firstly, for the Sierpinski fractal structure, disorders are introduced by add columns to change the coupling strength. For the inner edge state, two cylinders are added at each corner of the inner boundary (the same with that shown in Fig.C2(b)). For the inner corner state, add two columns to the diagonal corners respectively at the inner corner (the same with that shown in the inset of Fig.C2(e)). Figs. D1(b) and (d) show the displacement distributions of inner edge and corner states with disorders obtained by simulations. It can be seen that compared to the results for those with no disorders (Figs. D1(a) and (c)), after the introductions of disorders in the G(2) Sierpinski fractal structure, the energy distributions of inner edge and inner corner states are not affected obviously, which indicate that the robustness of the topological inner edge and corner states exists.

For the rhombus fractal structure, disorders are introduced in the same way as those for the Sierpinski fractal structure, the displacement distributions are shown in Fig.D2. Figs. D2(b) and (d) show the displacement distributions of inner edge and corner states with disorders obtained by simulations. It can be seen that compared to the results for those with no disorders (Figs. D2(a) and (c)), after the introductions of disorders, the energy distributions are changed slightly. The results indicate that the topological inner edge and corner states are robust.

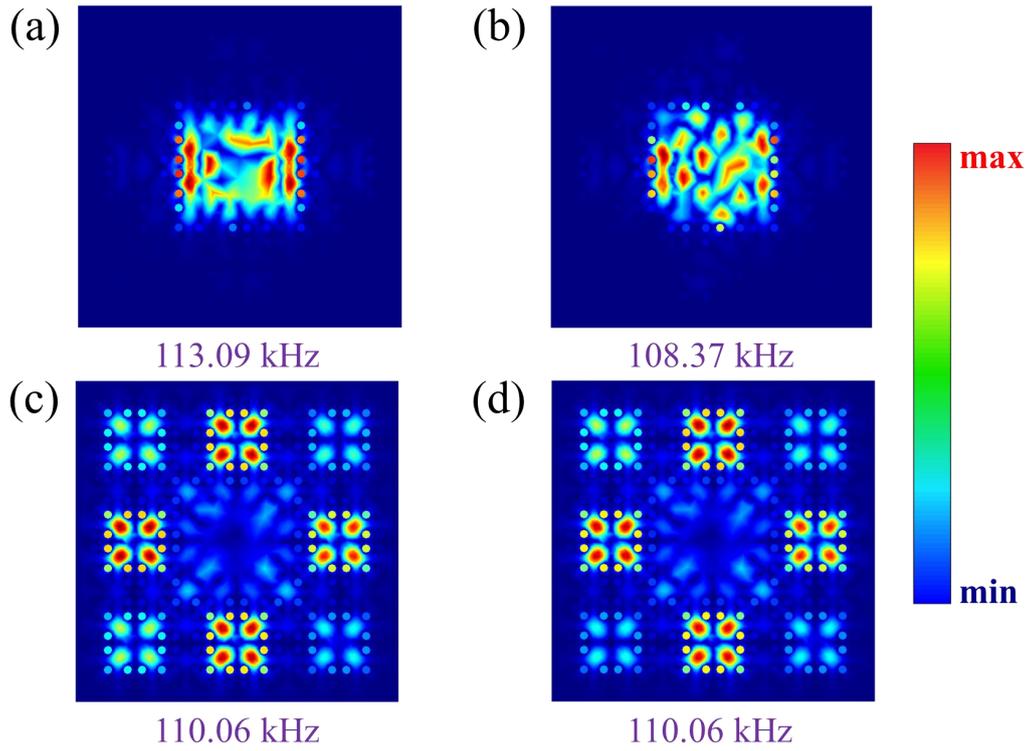

FIG. D1. Verification of the robustness of topological inner edge and corner states of the Sierpinski fractal structure. (a) Inner edge state with no defect. (b) Inner edge states with defects of disorder. (c) Inner corner state with no defect. (d) Inner corner states with defects of disorder.

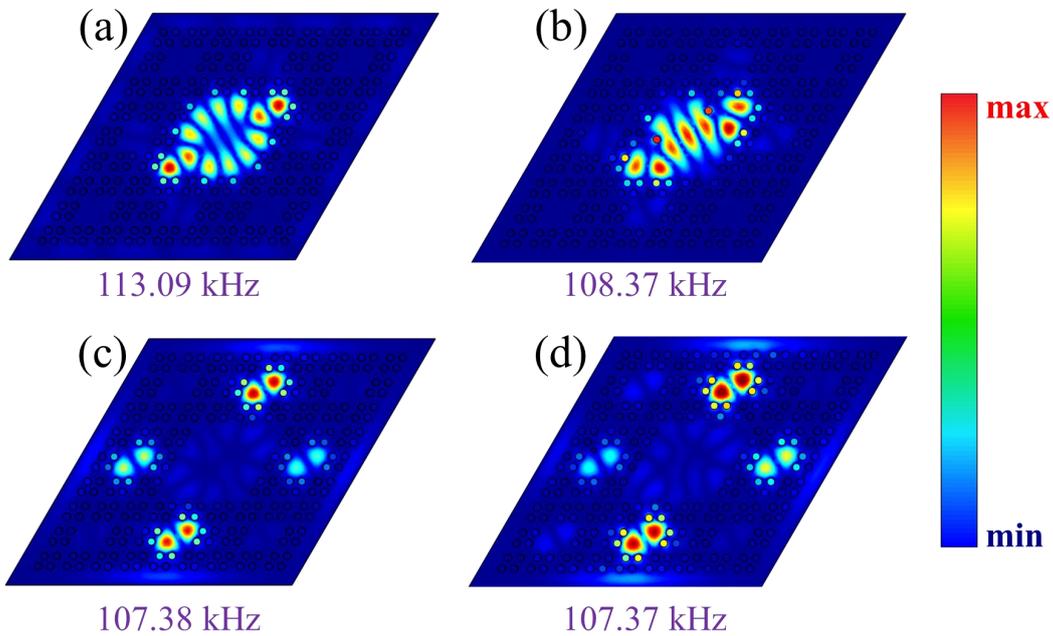

FIG. D2. Verification of the robustness of topological inner edge and corner states of rhombus fractal structures. (a) Inner edge states with no defect. (b) Inner edge states with defects of disorder. (c) Inner corner states with no defect. (d) Inner corner states with defects of disorder.

Then, the robustness of topological outer edge states is verified for the two kind of elastic fractal structures. For the Sierpinski fractal structure, to set the defect, two columns near the right edge are deleted, as shown in the red oval boxes in Fig. D3(b). Disorders are introduced by moving two columns to left by 1.5 mm and upwards by 1mm to change the coupling strength, as shown in the inset of Fig. D3(c); Besides, two columns near the right edge are lengthened by 3 mm to introduce disorders, as shown in the red oval boxes in Fig. D3(d). Figs. D3(b)-(d) show that after the introductions of defects and disorders in the G(2) Sierpinski fractal structure, although there is a small fluctuation for the frequency of edge states, and a small amount of energy spread to the edge, most energy is still concentrated near the edge, which indicates that the introduction of defects and disorders have slight influences on the energy distributions of edge states. For the rhombic fractal structure, the setting of defects and disorders are same with those for the Sierpinski fractal structure, and the results are shown in Fig. D4. Similar phenomena can be observed. Thus, the topological edge states of Sierpinski and rhombic fractal structures are robust to defects and disorders.

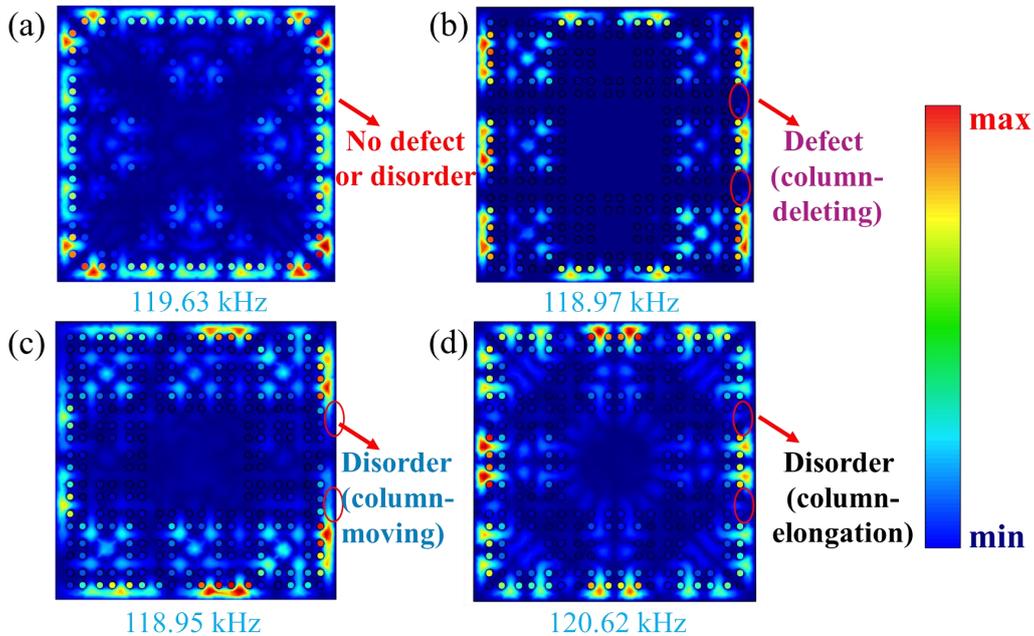

FIG. D3. Verification of the robustness of topological edge states of the Sierpinski fractal structure. (a) Model with no defect. (b) Perturbed model with defect (column-deleting). (c) Perturbed model with disorder (column-moving). (d) Perturbed model with disorder (column-elongation).

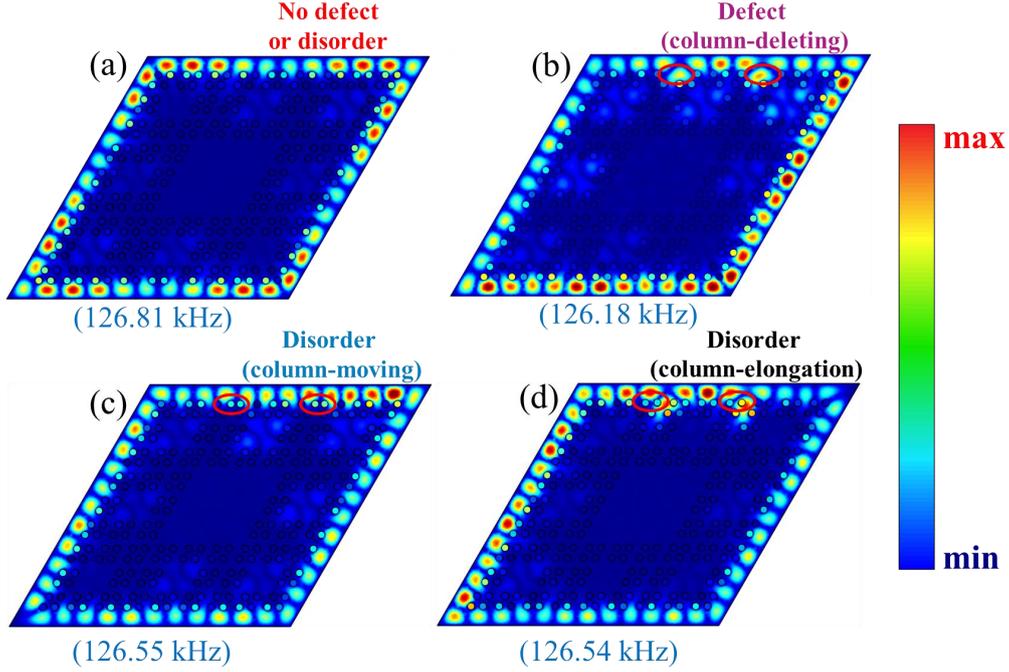

FIG. D4. Verification of the robustness of topological edge states of the rhombus fractal structure. (a) Model with no defect. (b) perturbed model with defect (column-deleting). (c) perturbed model with disorder (column-moving). (d) perturbed model with disorder (column-elongation).

## Appendix E. Simulated eigenstates of the G(3) fractal model

The eigenstates of the G(3) Sierpinski fractal structure are calculated. According to the calculation method in Appendix B, the total number of eigenstates in the G(3) generation structure is 2048, including 4 outer corner states, 584 inner corner states and 400 edge states. Fig. E1 shows the eigenstates of G(3) fractal structure obtained by numerical simulations. It is showed that there are 1060 bulk states (black), 584 inner corner states (green) and 400 edge states, which include 192 inner edge states (purple) and 208 outer edge states (blue). The four outer corner states are represented by red inverted triangles. Displacement distributions of the G(3) fractal structure are shown in Figs. E1(b)-(d), which include bulk state (82.233 kHz), inner edge state (112.76 kHz), inner corner state (97.043 kHz), outer edge state (121.03 kHz) and outer corner state (125.94 kHz). The energy of the inner edge, inner corner, outer edge, outer corner states maintain a high concentration, and the frequency range of the calculated states of the generation G(3) structure is the same as that of the generation G(2), which is due to the excellent self-similarity of the fractal structure.

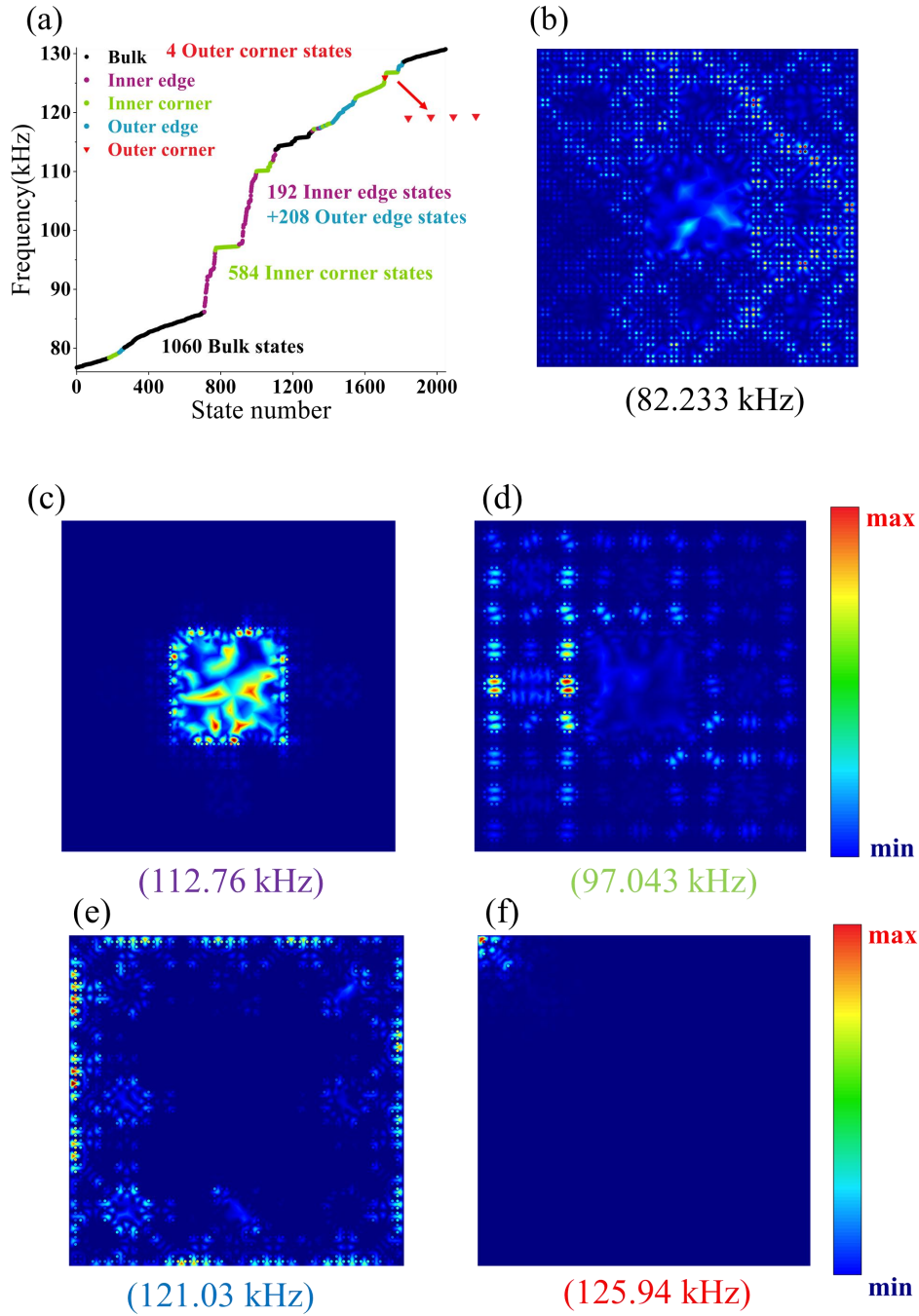

FIG. E1. (a) The eigen frequencies of the G(3) Sierpinski fractal plate at topological state obtained by simulations. Black, purple, green, blue and red dots denote the bulk states, inner edge states, inner corner states, outer edge states and outer corner states, respectively. (b)-(f) denote displacement distributions of the bulk state at 82.233 kHz, the inner edge state at 112.76 kHz, the inner corner state at 97.043 kHz, the outer edge state at 121.03 kHz and the outer corner state at

125.94 kHz, respectively.

Fig. E2 shows the eigenstates of the G(3) elastic rhombus fractal structure. In Fig. E2(a), the 2π/3 outer corner states, π/3 outer corner states, outer edge states, and inner edge states, inner corner states and bulk states are shown with different colors. Figs. E2(b-g) shows displacement distributions of the bulk state (85.217 kHz), inner edge state (94.862 kHz), inner corner state (112.77 kHz), and outer edge state (126.94 kHz), π/3 outer corner state (97.908 kHz) and 2π/3 outer corner state (115.83 kHz) obtained by simulations. It is shown that the energy of the corner and edge states remain highly concentrated.

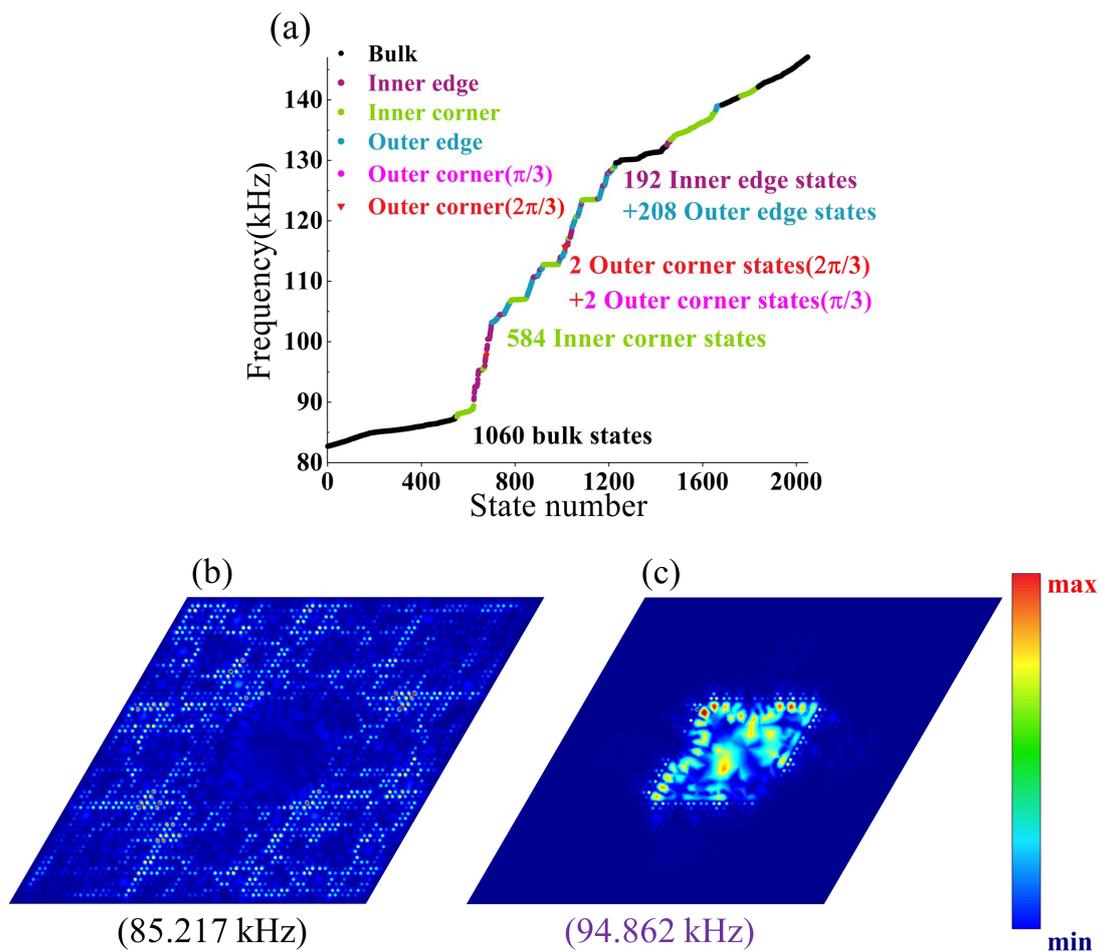

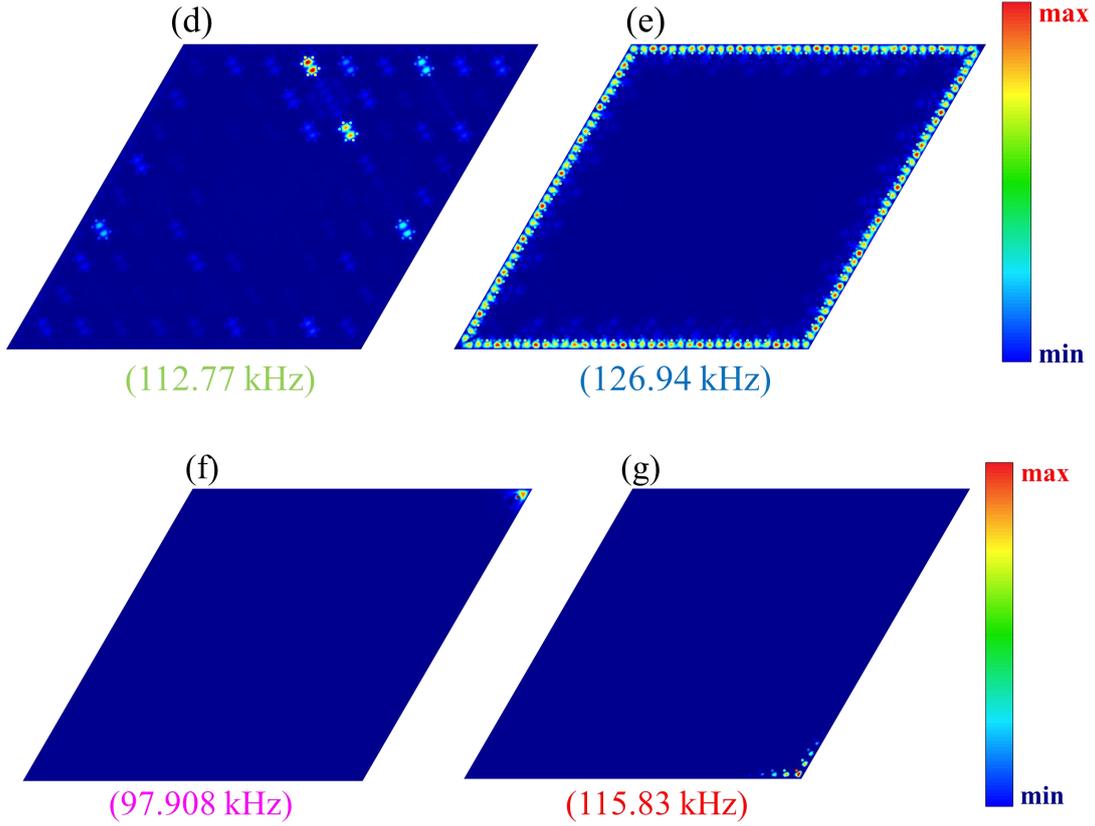

FIG. E2. (a) The eigen frequencies of the G(3) rhombus fractal plate at topological state obtained by simulations. Black, purple, green, blue, red and pink denote the bulk state, inner edge state, inner corner state, edge state, $2\pi/3$ outer corner state, $\pi/3$ outer corner states, respectively. (b)-(f) displacement distributions of the bulk state (85.217 kHz), inner edge state (94.862 kHz), inner corner state (112.77 kHz), edge state (126.94 kHz), $\pi/3$ outer corner state (97.908 kHz) and $2\pi/3$ outer corner state (115.83 kHz), respectively.

## Appendix F. Simulation of eigenstates of periodic elastic structures

In order to verify further the superiority of fractal structures compared to periodic structures in elastic system, the periodic square structure with an integer dimension is designed (Fig. F1(c)), and the coupling parameters (the intracellular coupling is greater than the extracellular coupling) and other parameters of the structure are consistent with the fractal structures. It is found that the periodic structure has 24 topological edge states, 4 topological corner states, and the rest are bulk states, as shown in Fig. F1(a). In the trivial state, only bulk states emerged, as shown in Fig. F1(b). Fig. F1(d)-(f) shows displacement distributions of the bulk state (82.629 kHz), outer edge state (120.54 kHz) and outer corner state (125.47 kHz), the frequency of which are close to those for the Sierpinski fractal structure. However, different from the fractal structure, there is no inner corner state and inner edge state in the periodic structure, and due to the disappearance of the self-similarity of the structure, the periodic structure has far fewer edge states than the fractal structure. The number of topological states in the fractal structure is 156, while the number of edge states in the integer dimension is only 28. Thus, the elastic fractal structure has a great advantage in the richness of topological states, compared with the periodic structure.

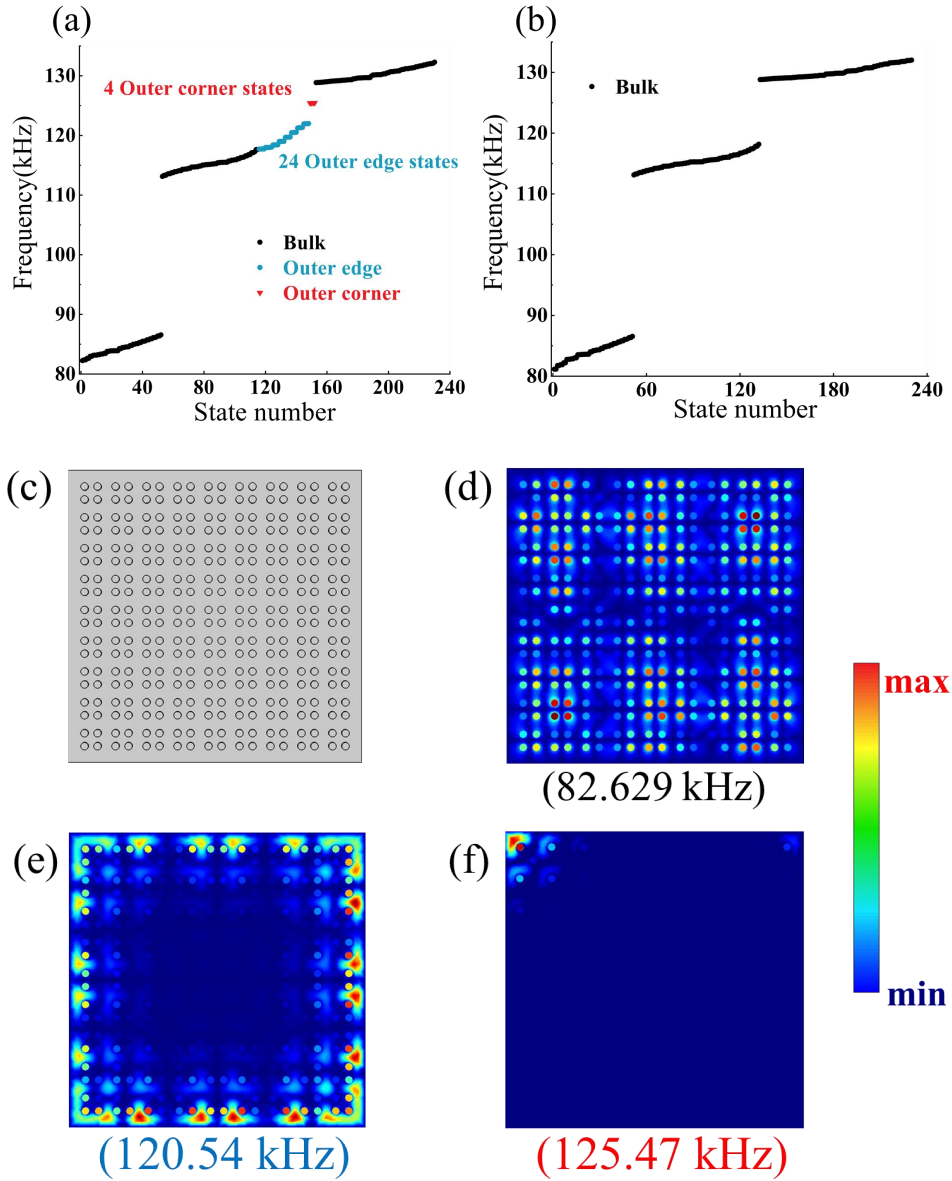

FIG. F1. The eigenstates of the periodic square structure obtained by simulations. Red inverted triangles, blue dots and black dots denote topological corner states, edge states and bulk states, respectively. (a) Eigen frequencies of the periodic square structure under the coupling parameters where topology occurs. (b) Eigen frequencies of the periodic square structure under the coupling

parameters where no topology occurs. (c) The designed periodic square structure. (d)-(f) Displacement distributions of the bulk state at 82.629 kHz, the edge states at 120.54 kHz and the outer corner state at 125.47 kHz.

## Appendix G. Topology of $\pi/3$ and $2\pi/3$ outer corner states in rhombus plates

A topological index N [58,59] is introduced to differentiate topological and trivial corner states in rhombus plates. This index captures the interaction between the bulk Hamiltonian topology and the defect topology. It can be seen that at the $\pi/3$ corner in the rhombus structure, $N_+ = 2$ and $N_- = 2$, the topological index $N = 0$. At the $2\pi/3$ corner, $N_+ = 1$ and $N_- = 2$, the topological index $N = 1$, which indicates the occurrence of the stable mode, namely, topological protection is generated.

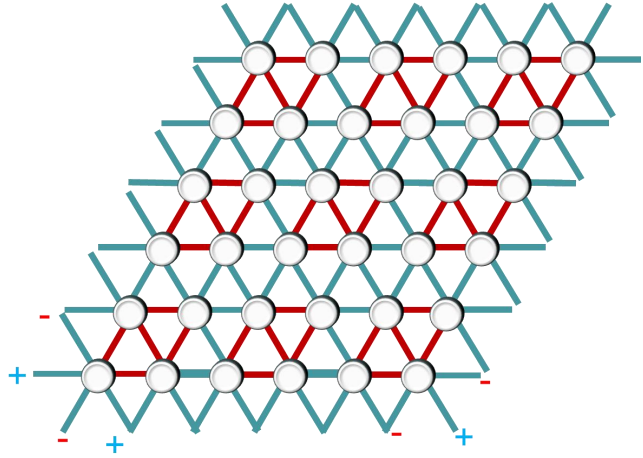

FIG. G1. Topology of Corner modes at $\pi/3$ and $2\pi/3$ corners in the rhombus lattice.